\documentclass[]{article}
\usepackage{graphicx}
\usepackage[colorlinks=true,breaklinks=true,citecolor=blue,linkcolor=blue,urlcolor=blue]{hyperref}
\usepackage{lscape}

\usepackage{amssymb}
\usepackage{amsmath}

\title{\textsl{\textbf{Study \textit{Pelamis} system to capture energy of ocean wave}}}

\author{\textbf{Ricardo Gobato$^{1}$}\\ Secretaria de Estado da Educa\c{c}\~ao do Paran\'{a} (SEED/PR),\\ Av. Maring\'{a}, 290, Jardim Dom Bosco,\\ Londrina/PR, 86060-000, Brasil\\ \\
	\textbf{Alekssander Gobato$^{2}$}\\ Faculdade Pit\'{a}goras Londrina, \\Rua Edwy Taques de Ara\'{u}jo, 1100,\\ Gleba Palhano, Londrina/PR, 86047-500, Brasil\\ \\
	 	 \textbf{Desire Francine Gobato Fedrigo$^{3}$}\\  Aeronautical Engineering Consulting\\ Consultant in processes LOA/PBN RNAV, Rua Lu\'{i}sa, 388s, ap. 05,\\ Vila Portuguesa, Tangar\'{a} da Serra/MT, 78300-000, Brasil\\ \\
	 \textbf{Corresponding authors}: $^{1}$ricardogobato@seed.pr.gov.br;\\ $^{2}$alekssandergobato@hotmail.com; $^{3}$desirefg@bol.com.br}

\begin{document}

\maketitle
 

\begin{abstract}
Over the years, energy has become vital for humans, enabling us to comfort, leisure, mobility and other factors. The quest for cheap energy sources, renewable and clean has grown in recent years, mainly for the reduction of effects that comes degrading nature, allowing scientists and engineers to search for new technologies. Many energy sources have been researched for proper funding where some stand out for their ease of obtaining, by other low cost and others by being renewable. The main objective of this work is to study one of these energy sources - wave energy, whose capture is still in development. This energy comes from the waves of the sea and is 100\% renewable and with minimal environmental impact when compared to hydro, nuclear, coal, thermal, etc. The system studied here is the \textit{Pelamis} system. 
\end{abstract}

\section{Introduction}
The main objective of this work is to study the wave energy, whose capture is still in development. This energy comes from the waves of the sea and is 100\% renewable, clean and without significant environmental impact when compared to hydro, nuclear, coal, thermal, etc. The system studied here is the \textit{Pelamis} system.\

Many energy sources have been researched for proper funding where some stand out for their ease of obtaining, by other low cost and others by being renewable. Here we study one of these energy sources - tidal, whose capture is still in development. This energy comes from the waves of the sea and is 100\% renewable, and the system dealt with here is the \textit{Pelamis} system. Over the years, energy has become vital for humans, enabling us to comfort, leisure, mobility and other factors. The search for cheap and renewable energy sources has grown in recent years, mainly for the reduction of effects that is degrading nature, allowing scientists and engineers to search for new technologies. Some countries where there are not many ways to capture energy, are seeking alternative sources such as wind, solar, thermal, tidal and many others. Among the energy harvesting methods mentioned above, a project created by Chinese in 1988 \cite{HeHongzhous2013}, in order to be a source of endless and totally clean energy that caught our attention. It is a system that operates at sea generating energy through the movement of the waves, this project is called Pelamis.\

In Europe, during the early stages of testing full-scale between 2004 and 2007, analyzed the "size" of technology and how good the project could be and this generated a lot of other ideas for modeling. The prototype had 120 meters long and 3.5 meters wide, which were connected four tubes that were the Pelamis floating and three converter modules of energy.
In 2008 after many tests, other equipment were installed three Pelamis the coast of Peniche, Portugal. But unfortunately the project was terminated earlier than planned because the Babcock \& Brown company, who managed the project went bankrupt. It ended what was called Pelamis P1, a company called ScottishPower Renewables, took over the project and currently the company is developing the so-called P2 Pelamis, which has somewhere around 180 meters long and four meters wide with about 1350 tons, which were added over a buoy, a total of five, and one power converter module, making it more efficient than the Pelamis P1.\

\begin{figure}
	\begin{center}
		\includegraphics[scale=0.655]{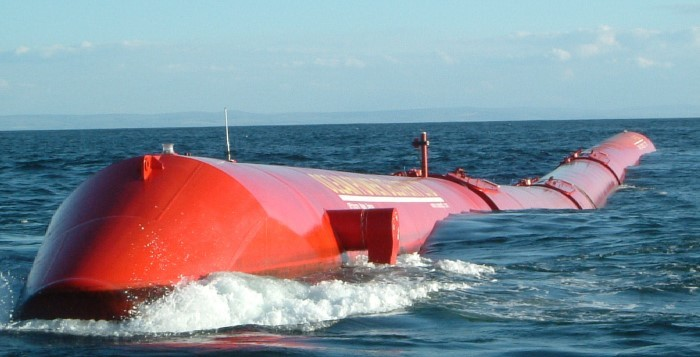}
		\caption{\small{Pelamis P1 operating in the coast town of Peniche, Portugal.\cite{Rodrigues}}}\label{fig:sweeps}
	\end{center}
\end{figure}

\begin{figure}
	\begin{center}
		\includegraphics[scale=0.38]{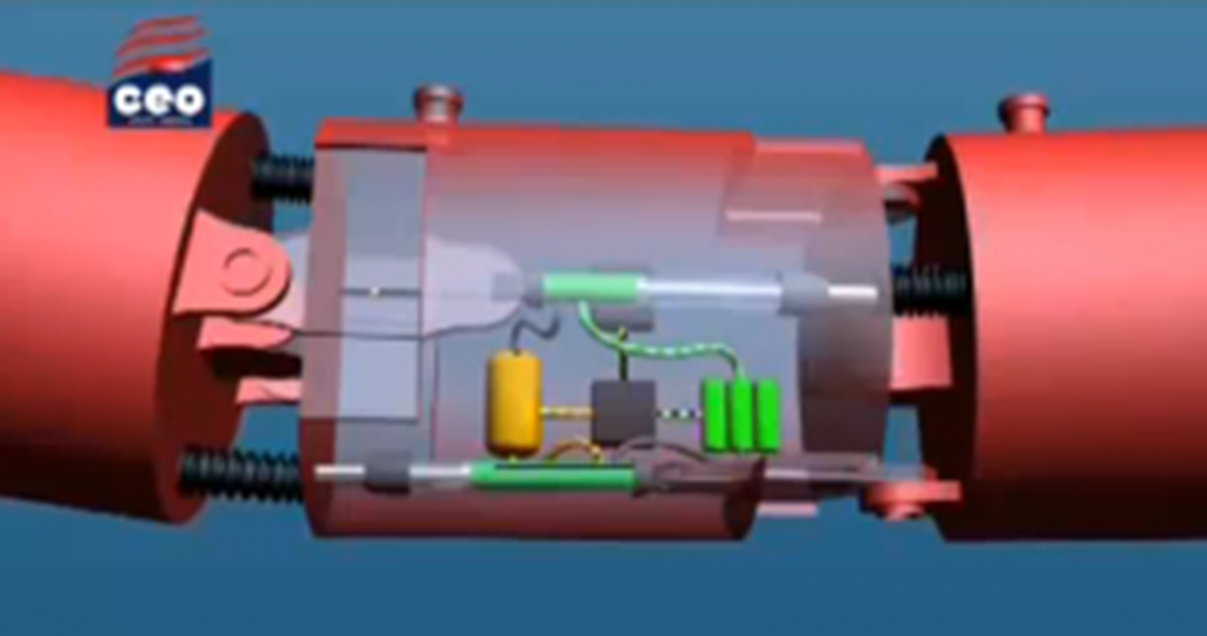}
		\caption{\small{Cylinders (between two buoys) pump oil for a motor. \cite{Harrington2014}}}\label{fig:sweeps}
	\end{center}
\end{figure}

However, like any new project, this created many difficulties for its researchers. Thus, the aim of this paper is to study the method of Pelamis wave energy capture. To achieve this goal we will analyze its operation and how effective it can be, compared to other renewable energy sources used today. \cite{HeHongzhous2013}\

\section{Gear}
The \textit{Pelamis} system behaves essentially as a large float of approximately 130 meters long and 3.5 meters in diameter which is floating horizontally and transverse to the waves. In operation, it moves along with the waves, and this movement causes it to pump oil for a generator which is on the rear Pelamis, causing the generator produces energy that is channeled to a receiver which converts this energy into electrical energy , suitable for our use.\cite{HeHongzhous2013, RichardYemm2011}

\subsection{Pelamis WEC technology - Concept}

\textbf{1}. Articulated cylinder\\
\textbf{2}. Swings head-on to incident waves\\
\textbf{3}. 4 x main segments, 3 x joints\\
\textbf{4}. Wave induced joint motion resisted to absorb power\\
\textbf{5}. 140m long, 3.5m diameter\\
\textbf{6}. 750kW rated power\\
\textbf{7}. Capacity factor 0.25-0.4\\

\begin{figure}
	\begin{center}
		\includegraphics[scale=0.38]{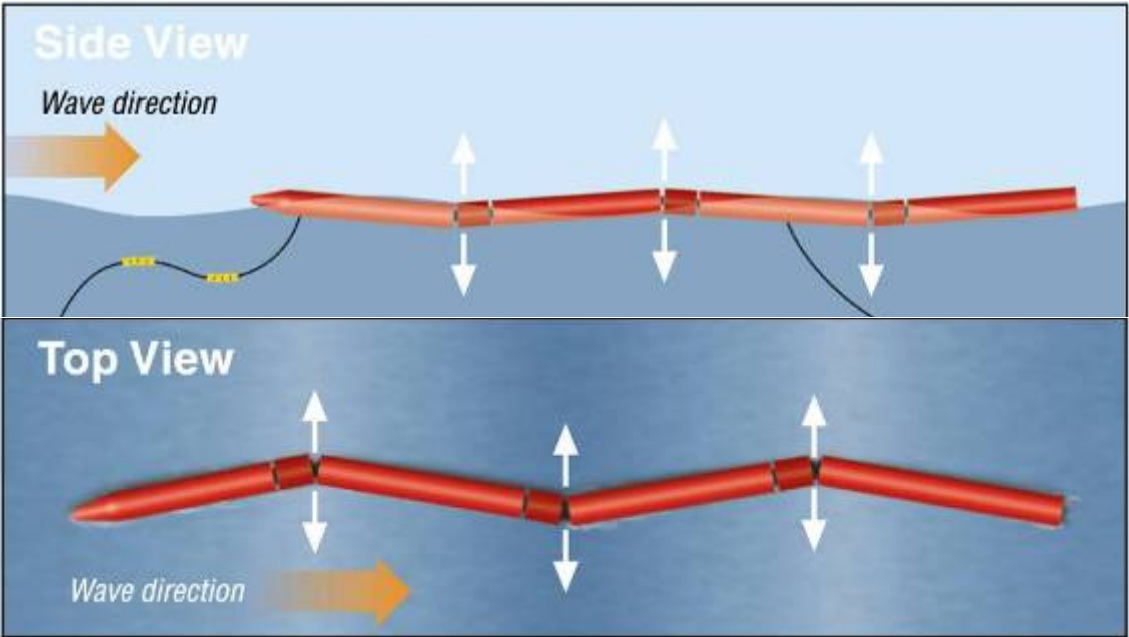}
		\caption{\small{Cylinders (between two buoys) pump oil for a motor. Articulated cylinder. Swings head-on to incident waves. \cite{Carcas}}}\label{fig:sweeps}
	\end{center}
\end{figure}

The Power Train is Developer of UK’s first offshore and Europe’s largest wind-farm (700MW) in Western Isles, is the major international utility UK’s largest wind developer, Amec, Scottish Power, Figure (5). \cite{Carcas}\\

A Portuguese energy company called Enersis funded a commercial wave energy project in Northern Portugal. Construction began at the end of October 2006. The project used \textit{Pelamis} system.

\begin{figure}
	\begin{center}
		\includegraphics[scale=0.4]{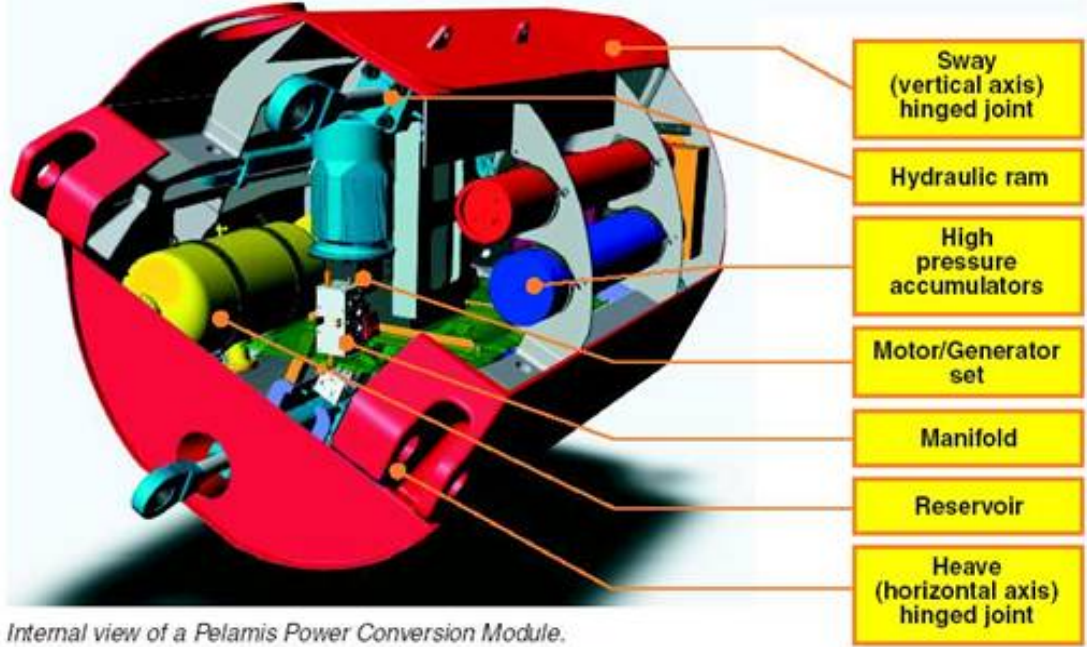}
		\caption{\small{Pelamis WEC technology – Power Train. Cylinders pump oil for a motor. \cite{Carcas}}}\label{fig:sweeps}
	\end{center}
\end{figure}

\begin{figure}
	\begin{center}
		\includegraphics[scale=0.5]{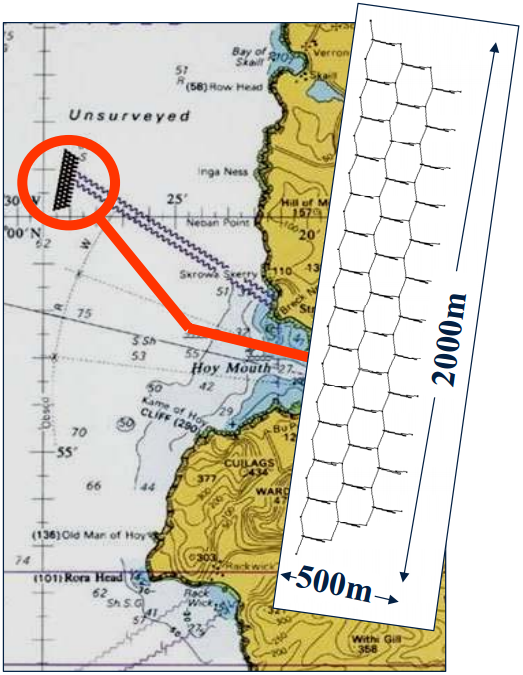}
		\caption{\small{Developer of UK’s first offshore and
				Europe’s largest wind-farm (700MW)
				in Western Isles. Consortium development
				- Plan: phased 22.5MW wave farm
				- Grid connection applied for
				- Target: first stage 2005/6
				- Key DTI/Scottish Executive market
				enablement mechanisms. \cite{Carcas}}}\label{fig:sweeps}
	\end{center}
\end{figure}

\section{Wave Power}
Wave power is the transport of energy by ocean surface waves, and the capture of that energy to do useful work – for example, electricity generation, water desalination, or the pumping of water (into reservoirs). A machine able to exploit wave power is generally known as a wave energy converter (WEC). \cite{Wavepower} 

Wave power is distinct from the diurnal flux of tidal power and the steady gyre of ocean currents. Wave-power generation is not currently a widely employed commercial technology, although there have been attempts to use it since at least 1890 \cite{Miller2004}. In 2008, the first experimental wave farm was opened in Portugal, at the Aguçadoura Wave Park \cite{Lim2008}. The major competitor of wave power is offshore wind power, with more visual impact. \cite{Wavepower} \

\subsection{Physical concepts}
Waves are generated by wind passing over the surface of the sea. As long as the waves propagate slower than the wind speed just above the waves, there is an energy transfer from the wind to the waves. Both air pressure differences between the upwind and the lee side of a wave crest, as well as friction on the water surface by the wind, making the water go into the shear stress causes the growth of the waves. \cite{Phillips1977}

\begin{figure}
	\begin{center}
		\includegraphics[scale=0.65]{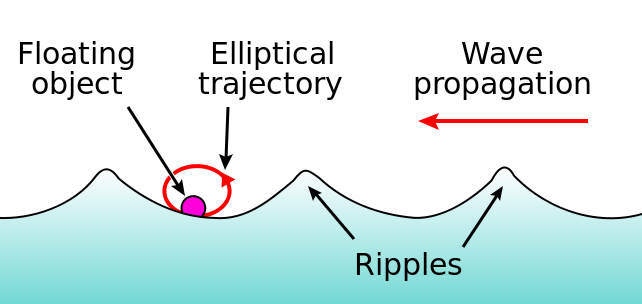}
		\caption{\small{When an object bobs up and down on a ripple in a pond, it experiences an elliptical trajectory. \cite{Wavepower}}}\label{fig:sweeps}
	\end{center}
\end{figure}

Wave height is determined by wind speed, the duration of time the wind has been blowing, fetch (the distance over which the wind excites the waves) and by the depth and topography of the seafloor (which can focus or disperse the energy of the waves). A given wind speed has a matching practical limit over which time or distance will not produce larger waves. When this limit has been reached the sea is said to be ``fully developed".\

In general, larger waves are more powerful but wave power is also determined by wave speed, wavelength, and water density.\

\begin{figure}
	\begin{center}
		\includegraphics[scale=0.65]{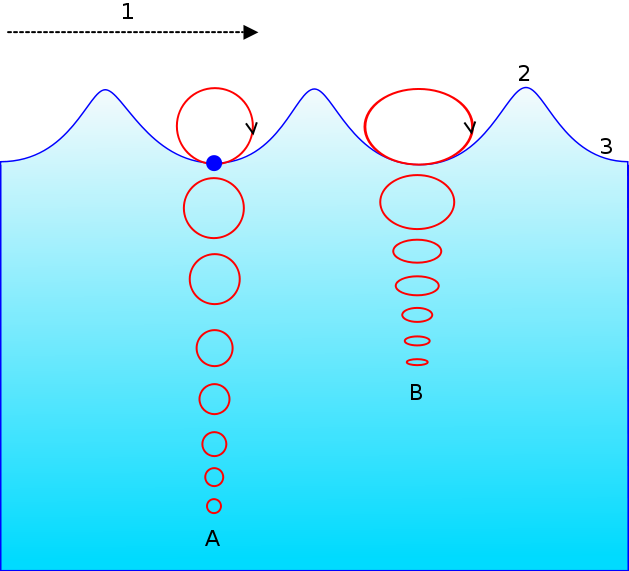}
		\caption{\small{Motion of a particle in an ocean wave. A = At deep water. The orbital motion of fluid particles decreases rapidly with increasing depth below the surface. B = At shallow water (ocean floor is now at B). The elliptical movement of a fluid particle flattens with decreasing depth.	1 = Propagation direction. 2 = Wave crest.		3 = Wave trough. \cite{Wavepower}}}\label{fig:sweeps}
		\end{center}
\end{figure}

\begin{figure}
	\begin{center}
		\includegraphics[scale=0.65]{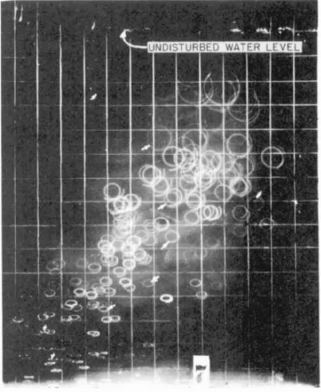}
		\caption{\small{Photograph of the water particle orbits under a – progressive and periodic – surface gravity wave in a wave flume. The wave conditions are: mean water depth d = 2.50 ft (0.76 m), wave height H = 0.339 ft (0.103 m), wavelength $\lambda$ = 6.42 ft (1.96 m), period T = 1.12 s. \cite{Wiegel1950, Herbich2000}}}\label{fig:sweeps}
	\end{center}
\end{figure}

Oscillatory motion is highest at the surface and diminishes exponentially with depth. However, for standing waves (clapotis) near a reflecting coast, wave energy is also present as pressure oscillations at great depth, producing microseisms \cite{Phillips1977}. These pressure fluctuations at greater depth are too small to be interesting from the point of view of wave power.\

The waves propagate on the ocean surface, and the wave energy is also transported horizontally with the group velocity. The mean transport rate of the wave energy through a vertical plane of unit width, parallel to a wave crest, is called the wave energy flux (or wave power, which must not be confused with the actual power generated by a wave power device).

\subsection{Wave power equation}

In deep water where the water depth is larger than half the wavelength, the wave energy flux is
\begin{equation}
\
P=\frac{\rho{}g^2}{64\pi{}}H_{m0}^2T_e\approx{}\left(0.5\frac{kW}{m^3s}H_{m0}^2\right)T_e
\
\end{equation}

with P the wave energy flux per unit of wave-crest length, H$_{m0}$ the significant wave height, Te the wave energy period, ρ the water density and g the acceleration by gravity. The above equation states that wave power is proportional to the wave energy period and to the square of the wave height. When the significant wave height is given in metres, and the wave period in seconds, the result is the wave power in kilowatts (kW) per metre of wavefront length. \cite{M.J.Tucker2001, Strathclyde, United, Scotland}\

The energy flux is\ 

\begin{equation}
\
P=\frac{1}{16}H_{m0}^2c_g
\
\end{equation}

the group velocity see \cite{Herbich2000}. The group velocity is\
\begin{equation}
\ 
c_g=\tfrac{g}{4\pi}T 
\
\end{equation}

see the collapsed table "Properties of gravity waves on the surface of deep water, shallow water and at intermediate depth, according to linear wave theory.\

Example: Consider moderate ocean swells, in deep water, a few km off a coastline, with a wave height of 3 m and a wave energy period of 8 seconds. Using the equation to solve for power, we get\

\begin{equation}
\
E=\left(0.5\frac{kW}{m^3s}\right)\left(3.m^2\right)\left(8.s\right)\approx{}12\frac{kW}{m}
\
\end{equation}

meaning there are 12 kilowatts of power potential per meter of wave crest.\

In major storms, the largest waves offshore are about 15 meters high and have a period of about 15 seconds. According to the above equation, such waves carry about 112.5 kW of power across each metre of wavefront. An effective wave power device captures as much as possible of the wave energy flux. As a result, the waves will be of lower height in the region behind the wave power device.\

\subsection{Wave energy and wave-energy flux}
In a sea state, the average(mean) energy density per unit area of gravity waves on the water surface is proportional to the wave height squared, according to linear wave theory:  \cite{Phillips1977, God2000, Holthuijsen2007}\

\begin{equation}
\
E=\frac{1}{8}\rho{}gH_{m0}^2
\
\end{equation}

where E is the mean wave energy density per unit horizontal area (J/m$^{2}$), the sum of kinetic and potential energy density per unit horizontal area. The potential energy density is equal to the kinetic energy \cite{Phillips1977}, both contributing half to the wave energy density E, as can be expected from the equipartition theorem. In ocean waves, surface tension effects are negligible for wavelengths above a few decimetres.\

Here, the factor for random waves is $1⁄16$, as opposed to $1⁄8$ for periodic waves – as explained hereafter. For a small-amplitude sinusoidal wave\

\begin{equation}
\
\eta=a\,\cos\, 2\pi\left(\frac{x}{\lambda}-\frac{t}{T}\right) 
\
\end{equation}
 
with wave amplitude \textit{a}, the wave energy density per unit horizontal area is\

\begin{equation}
\  
E=\frac{1}{2}\rho g a^2
\
\end{equation}

or

\begin{equation}
\  
 E=\frac{1}{8}\rho g H^2
\
\end{equation}

using the wave height 
$H\,=\,2\,a\,$ for sinusoidal waves. In terms of the variance of the surface elevation\

\begin{equation}
\
m_0=\sigma_\eta^2=\overline{(\eta-\bar\eta)^2}=\frac{1}{2}a^2
\
\end{equation}

the energy density is $E=\rho g m_0\,$. Turning to random waves, the last formulation of the wave energy equation in terms of $m_0\,$ is also valid \cite{Holthuijsen2007a}, due to \textit{Parseval's theorem}. Further, the significant wave height is defined as $H_{m0}=4\sqrt{m_0},$ leading to the factor $1/16$ in the wave energy density per unit horizontal area.\

As the waves propagate, their energy is transported. The energy transport velocity is the group velocity. As a result, the wave energy flux, through a vertical plane of unit width perpendicular to the wave propagation direction, is equal to: \cite{Phillips1977, Reynolds1877}

\begin{equation}
\
P=Ec_{g}
\
\end{equation}

with c$_{g}$ the group velocity (m/s). Due to the dispersion relation for water waves under the action of gravity, the group velocity depends on the wavelength $\lambda$, or equivalently, on the wave period \textit{T}. Further, the dispersion relation is a function of the water depth \textit{h}. As a result, the group velocity behaves differently in the limits of deep and shallow water, and at intermediate depths:\ \cite{Phillips1977, Scotland} \

Properties of gravity waves on the surface of deep water, shallow water and at intermediate depth, according to linear wave theory.

\subsection{Significant wave height}
In physical oceanography, the significant wave height (SWH or \textit{H$_{s}$}) is defined traditionally as the mean wave height (trough to crest) of the highest third of the waves \textit{(H$_{1/3}$)}. Nowadays it is usually defined as four times the standard deviation of the surface elevation – or equivalently as four times the square root of the zeroth-order moment (area) of the wave spectrum  \cite{Holthuijsen2007}. The symbol \textit{H$_{m0}$} is usually used for that latter definition. The significant wave height may thus refer to \textit{H$_{m0}$} or \textit{(H$_{1/3}$)}; the difference in magnitude between the two definitions is only a few percent. \cite{significantwave}\

The original definition resulted from work by the oceanographer Walter Munk during World War II \cite{Denny1988, Munk1944}. The significant wave height was intended to mathematically express the height estimated by a ``trained observer". It is commonly used as a measure of the height of ocean waves.

\begin{figure}
	\begin{center}
		\includegraphics[scale=0.46]{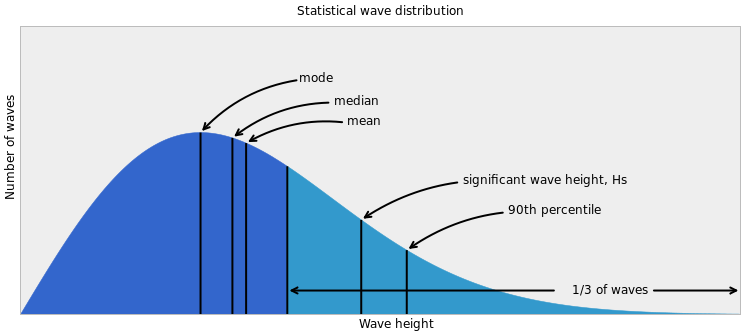}
		\caption{\small{Statistical distribution of ocean wave heights. \cite{significantwave}}}\label{fig:sweeps}
	\end{center}
\end{figure}

\begin{figure}
	\begin{center}
		\includegraphics[scale=0.65]{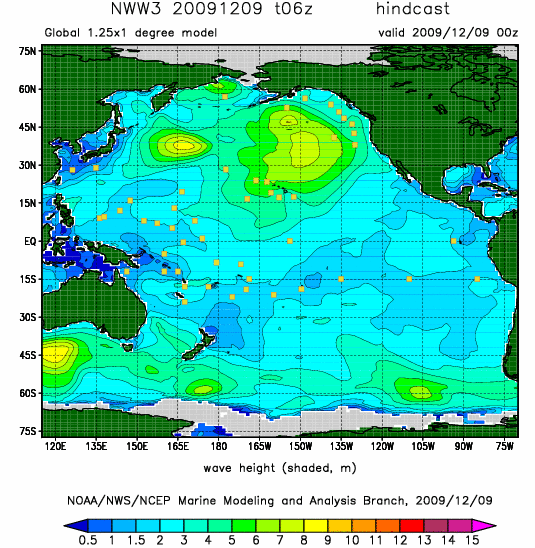}\\
		\includegraphics[scale=0.65]{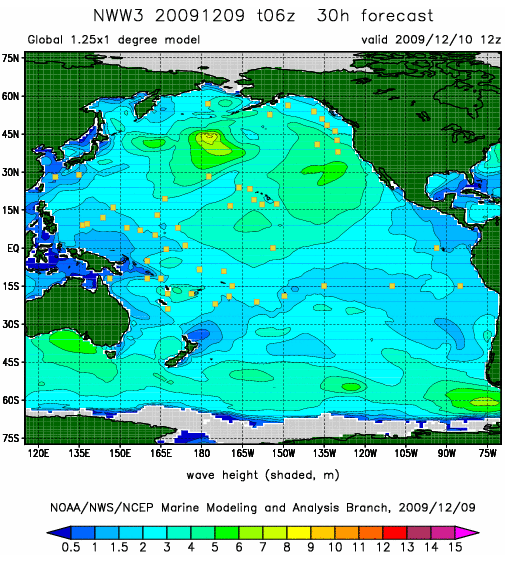}\\
		\caption{\small{NOAA WAVEWATCH III \textregistered 120-hour Forecast for the North Atlantic \cite{significantwave, NOAA}}}\label{fig:sweeps}
	\end{center}
\end{figure}

\begin{figure}
	\begin{center}
		\includegraphics[scale=0.65]{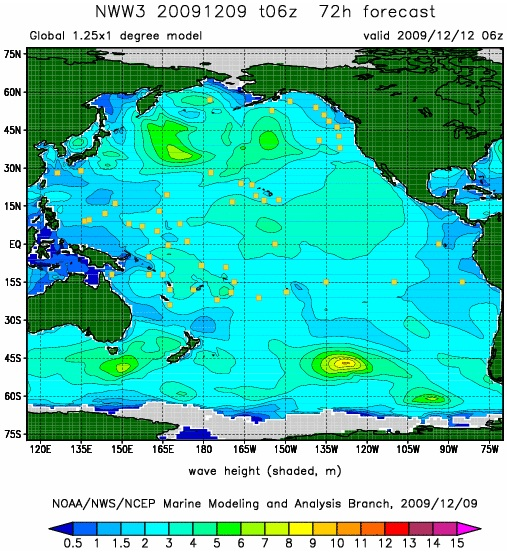}\\
		\includegraphics[scale=0.65]{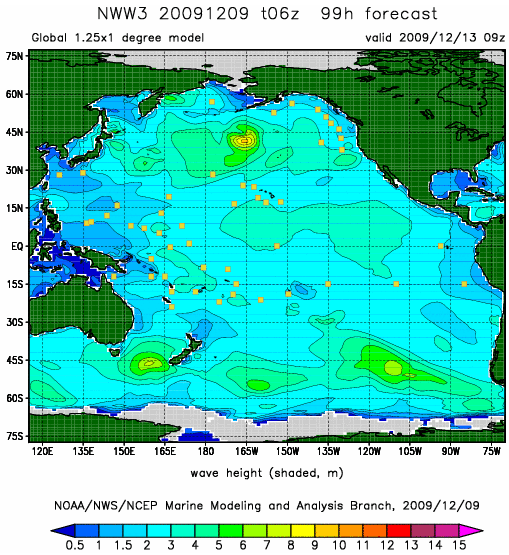}\\
		\caption{\small{NOAA WAVEWATCH III \textregistered 120-hour Forecast for the North Atlantic \cite{significantwave, NOAA}}}\label{fig:sweeps}
	\end{center}
\end{figure}

\begin{figure}
	\begin{center}
		\includegraphics[scale=0.65]{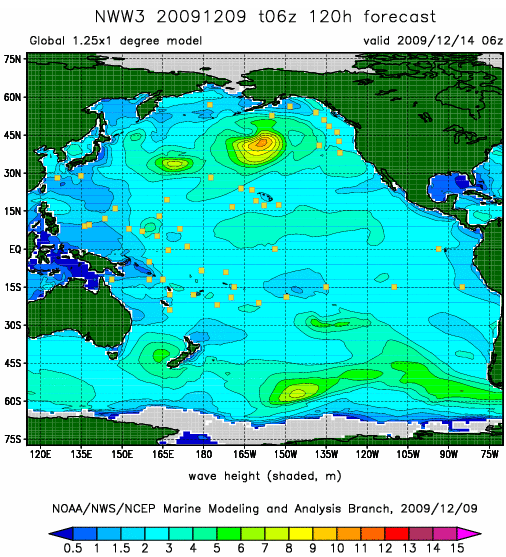}\\
		\includegraphics[scale=0.65]{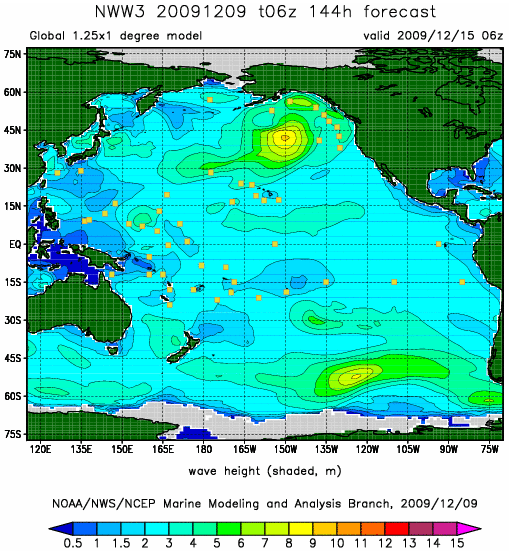}\\
		\caption{\small{NOAA WAVEWATCH III \textregistered 120-hour Forecast for the North Atlantic \cite{significantwave, NOAA}}}\label{fig:sweeps}
	\end{center}
\end{figure}

Significant wave height, scientifically represented as \textit{H$_{s}$} or \textit{H$_{sig}$}, is an important parameter for the statistical distribution of ocean waves. The most common waves are less in height than \textit{H$_{s}$}. This implies that encountering the significant wave is not too frequent. However, statistically, it is possible to encounter a wave that is much higher than the significant wave.\

Generally, the statistical distribution of the individual wave heights is well approximated by a Rayleigh distribution \cite{Ta}. For example, given that \textit{H$_{s}$} is 10 metres (33 feet), statistically:\

* 1 in 10 will be larger than 10.7 metres (35 ft)\

* 1 in 100 will be larger than 15.1 metres (50 ft)\

* 1 in 1000 will be larger than 18.6 metres (61 ft)\

This implies that one might encounter a wave that is roughly double the significant wave height. However, in rapidly changing conditions, the disparity between the significant wave height and the largest individual waves might be even larger.\

Although most measuring devices estimate the significant wave height from a wave spectrum, satellite radar altimeters are unique in measuring directly the significant wave height thanks to the different time of return from wave crests and troughs within the area illuminated by the radar. The maximum ever measured wave height from a satellite is 20.1m during a North Atlantic storm in 2011. \cite{JHana}\

\subsection{Weather forecasts}

The World Meteorological Organization stipulates that certain countries are responsible for providing weather forecasts for the world's oceans. These respective countries' meteorological offices are called Regional Specialized Meteorological Centers, or RSMCs. In their weather products, they give ocean wave height forecasts in significant wave height. In the United States, NOAA's National Weather Service is the RSMC for a portion of the North Atlantic, and a portion of the North Pacific. The Ocean Prediction Center and the Tropical Prediction Center's Tropical Analysis and Forecast Branch (TAFB) issue these forecasts.\

RSMCs use wind-wave models as tools to help predict the sea conditions. In the U.S., NOAA's WAVEWATCH III\textregistered  model is used heavily.\

A significant wave height is also defined similarly, from the wave spectrum, for the different systems that make up the sea. We then have a significant wave height for the wind-sea or for a particular swell. \cite{significantwave, NOAA}\

A sequence of wave motion pictures is described in Figures (10-12). \cite{NOAA}\

\subsection{Wind wave model}
In fluid dynamics, wind wave modeling describes the effort to depict the sea state and predict the evolution of the energy of wind waves using numerical techniques. These simulations consider atmospheric wind forcing, nonlinear wave interactions, and frictional dissipation, and they output statistics describing wave heights, periods, and propagation directions for regional seas or global oceans. Such wave hindcasts and wave forecasts are extremely important for commercial interests on the high seas \cite{Cox2002}. For example, the shipping industry requires guidance for operational planning and tactical seakeeping purposes. \cite{Cox2002}\

For the specific case of predicting wind wave statistics on the ocean, the term ocean surface wave model is used.\

Other applications, in particular coastal engineering, have led to the developments of wind wave models specifically designed for coastal applications. \cite{NOAA, windwave}\

The Figure (13) represents the direction of the wind waveat a given instant, for Wind wave model. \cite{NOAA}\

\begin{figure}
	\begin{center}
		\includegraphics[scale=0.65]{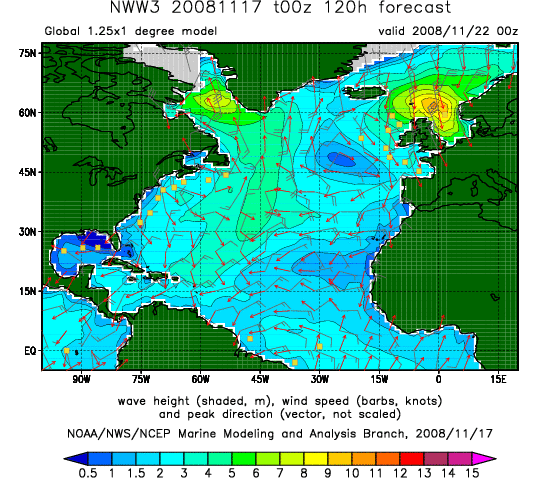}
		\caption{\small{Represents the direction of the wind waveat a given instant, for Wind wave model. NOAA WAVEWATCH III \textregistered Forecast for the North Atlantic. \cite{windwave, NOAA}}}\label{fig:sweeps}
	\end{center}
\end{figure}

\section{Fundamentals of Fluid Mechanics}

\subsection{Stress Tensor in a Moving Fluid}

We have seen that in a static fluid the stress tensor takes the form

\begin{equation}
\
{\sigma{}}_{ij}=-p{\delta{}}_{ij}
\
\end{equation}

where \textit{p }= -$\sigma{}$\textit{{\scriptsize ii}}/3 is the static
pressure: \textit{i.e.}, minus the normal stress acting in any direction. Now, the normal stress at a given point in a moving fluid generally varies with direction: \textit{i.e.},
the principal stresses are not equal to one another.
\\
However, we can still define the mean principal stress

\begin{equation}
{\frac{\left({\sigma{}}_{11}^{'}+{\sigma{}}_{22}^{'}+{\sigma{}}_{33}^{'}\right)}{3}=-\frac{{\sigma{}}_{ij}}{3}}\
\end{equation}

Moreover, given that the principal stresses are actually normal stresses (in a coordinate frame aligned with the principal axes), we can also regard $\sigma{}$\textit{{\scriptsize ii}}/3 as the mean normal stress. It is convenient to define pressure in a moving fluid as minus the \textit{mean normal stress}:
\textit{i.e.},

\begin{equation}
\
-p=\frac{1}{3}{\sigma{}}_{ii}
\
\end{equation}

where

\begin{equation}
\
d_{ii=}0
\
\end{equation}

Moreover, since $\sigma{}_{ij}$ and $\delta{}_{ij}$ are both symmetric tensors, it follows that $d_{ij}$ is also symmetric: \textit{i.e.},\

\begin{equation}
\
d_{ji}=d_{ij}
\
\end{equation}

where

\begin{equation}
\
e_{ij=}\frac{1}{2}\left(\frac{{\partial{}}_{{\upsilon{}}_i}}{{\partial{}}_{x_j}}+\frac{{\partial{}}_{{\upsilon{}}_j}}{{\partial{}}_{x_i}}\right)
\
\end{equation}

is called the rate of strain tensor. Finally, according to Equation (4), $_{dij}$ is a traceless tensor, which yields $3\alpha=-2\beta$, and

\begin{equation}
\
d_{ij}=2\mu{}\left(e_{ij}-\frac{1}{3}e_{kk}{\sigma{}}_{ij}\right)
\
\end{equation}

where $\mu=\beta$. We, thus, conclude that the most general expression for the stress tensor in an isotropic Newtonian fluid
is

\begin{equation}
\
{\sigma{}}_{ij}=-p{\delta{}}_{ij}+2\mu{}\left(e_{ij}-\frac{1}{3}e_{kk}{\sigma{}}_{ij}\right)
\
\end{equation}

where \textit{p}(\textbf{r},t) and $\mu(\textbf{r},t)$ are arbitrary scalars.\

the equation 

\begin{equation}
\frac{D_{{\nu{}}_i}}{D_t}=\frac{F_i}{\rho{}}+\frac{1}{\rho{}}\frac{{\partial{}}_{{\sigma{}}_i}}{{\partial{}}_{x_j}}
\
\end{equation}
\cite{Fitzpatrick}
\subsection{\textit{Navier-Stokes Equation}}
Equations (6), (8), and (9) can be combined to give the equation of motion of an isotropic, Newtonian,
classical fluid: \textit{i.e.},

\begin{equation}
\
\rho{}\frac{D_{{\nu{}}_i}}{D_t}=F_i\frac{{\partial{}}_p}{{\partial{}}_{x_i}}+\frac{\partial{}}{{\partial{}}_{x_j}}\left[\mu{}\left(\frac{{\partial{}}_{{\upsilon{}}_i}}{{\partial{}}_{x_j}}+\frac{{\partial{}}_{{\upsilon{}}_j}}{{\partial{}}_{x_i}}\right)\right]-\frac{\partial{}}{{\partial{}}_{x_i}}\left(-\frac{2}{3}\mu{}\frac{{\partial{}}_{{\upsilon{}}_j}}{{\partial{}}_{x_j}}\right)
\
\end{equation}

This equation is generally known as the\textit{ Navier-Stokes equation}. Now, in situations in which there are no strong
temperature gradients in the fluid, it is a good approximation to treat viscosity as a spatially uniform quantity, in which case the \textit{Navier-Stokes equation}  simplifies somewhat to give

\begin{equation}
\
\rho{}\frac{D_{{\nu{}}_i}}{D_t}=F_i\frac{{\partial{}}_p}{{\partial{}}_{x_i}}+\frac{\partial{}}{{\partial{}}_{x_j}}\left[\mu{}\left(\frac{{\partial{}}_{{\upsilon{}}_i}^2}{{{\partial{}}_{x_j}\partial{}}_{x_j}}+\frac{1}{3}\frac{{\partial{}}_{{\upsilon{}}_j}^2}{{\partial{}}_{x_i}{\partial{}}_{x_j}}\right)\right]
\
\end{equation}

When expressed in vector form, the above expression becomes
\begin{equation}
\
\rho{}\frac{D_V}{D_t}=\rho{}\frac{{\partial{}}_V}{{\partial{}}_t}+\left(v.\nabla{}\right)v=F-\nabla{}p+\mu{}\left[{\nabla{}}^2v+\frac{1}{3}\nabla{}\left(\nabla{}.v\right)\right]
\
\end{equation}
\cite{Fitzpatrick}

\section{Operation of Pelamis}
The Pelamis project is a way to obtain energy with the movement of the waves \cite{RichardYemm2011}. It fluctuates with the movement of waves and drives a series of floats, which consequently move a cylinder which pumps oil to a generator that produces hydraulic energy (Figures 1 and 2). After the generator into operation, a cable transmits the power generated to the surface and shortly thereafter it is converted into own electricity for our use, anywhere from 750 kW (based Pelamis P1 and normal sea conditions) . The interesting than its mode of operation is that the waves around you make more effect on energy production than the waves that pass through the front, it happens by the fact that the machine is in contact with all the water around you not only with the waves that pass through it. \cite{HeHongzhous2013, B.Drew2009}\

The system is basically comprised of hydraulic cylinders, pressure accumulators, motors / generators and tanks, where some parts are assembled into the tube making floating Pelamis - float \cite{HeHongzhous2013, RichardYemm2011}. The hydraulic cylinders pumping a fluid to a high pressure accumulator which smooths the movement performed by the waves, this system permits the maximum utilization when the waves are small, and also minimizes response even in storm. After the fluid exits the accumulator and is brought to an engine that produces power, the rate at which fluid enters the engine to produce energy, depends on the accumulator, which frees up more or less fluid, depending on sea conditions, which consequently makes it produces more or less energy. All sections (three in the case of P1 and four in the case of P2) are connected via a single cable that carries the energy produced to the coast. \cite{RichardYemm2011}

\section{Wave Energy}
The marine energy, compared to the others, is the most abundant source of energy we have \cite{Clement2002}. To get an idea of ​​how advantageous it is, we can compare it with other sources such as solar energy, which is achieved an average of 0.1- 0.3 KW (horizontally), while at sea on average 2 -3 kW (vertically). Another great advantage is minimal environmental impact when compared with solar energy, we need a large space for the installation of solar panels in the sea we do not do this, just install the generator, without generating any environmental impact. But, a disadvantage is that this technology is not yet well developed, making the forms is very limited uptake \cite{B.Drew2009}. Already the Hydro as large Itaipu, have their extremely high production (with 20 generating units and 14,000 MW of installed capacity, provides about 17\% of the energy consumed in Brazil and 75\% of the Paraguayan consumption), but it have high costs, both social, environmental and financial. The \textit{Pelamis} system is able to provide much smaller than a hydroelectric. It generates around 24 MW, but remembering that the project is still in development and can generate more power. The Pelamis does not damage the environment and does not bring social consequences, it is important to assess whether it is worth investing in it. \cite{HeHongzhous2013}

Limited negative environmental impact in use. Thorpe  \cite{Thorpe1999} details the potential impact and presents an estimation of the life cycle emissions of a typical nearshore device. In general, offshore devices have the lowest potential impact.\

Natural seasonal variability of wave energy, which follows the electricity demand in temperate climates \cite{Clement2002}.\

Waves can travel large distances with little energy loss. Storms on the western side of the Atlantic Ocean will travel to the western coast of Europe, supported by prevailing westerly winds.\

It is reported that wave power devices can generate power up to 90 per cent of the time, compared to $\sim$20-30 per cent for wind and solar power devices. \cite{B.Drew2009, Pelc2002, buoys}\

\subsection{The Pelamis wave energy converter}

The design of wave energy converters (WECs) has hitherto concentrated primarily on hydrodynamic efficiency \cite{Evans1985}. The Pelamis WEC is a promising new concept which is designed instead primarily for survival in extreme seas. This is accomplished by its end-on orientation to the waves, which enables the WEC negotiate breaking waves safely, and by its relatively small diameter (3.5m), which non-linearly limits power output in extreme conditions. \cite{Rainey}\

\begin{figure}
	\begin{center}
		\includegraphics[scale=0.5]{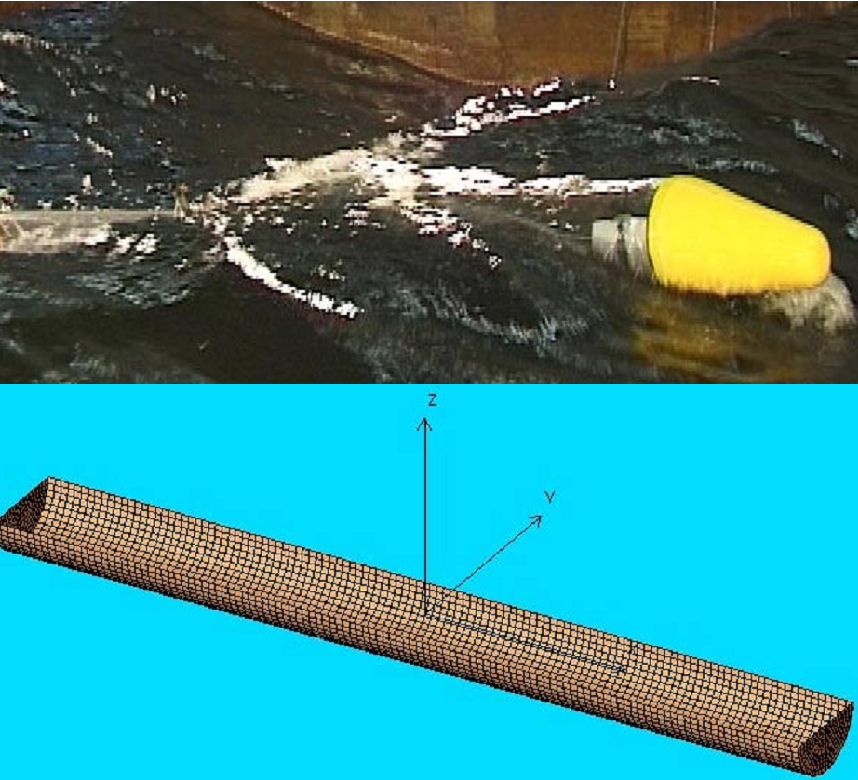}
		\caption{\small{The Pelamis wave energy converter. Analysis of a single segment with the diffraction program AQWA-LINE \cite{Newm1998}. The power take-off is from hydraulic jacks at the articulated joints. It has been found that the efficiency is greatly improved by making the transverse motions resonant (by suitable choice of transverse stiffness at the joints), and coupling them to the vertical motions \cite{Rainey}}}\label{fig:sweeps}
	\end{center}
\end{figure}

\subsection{The waves generated by a single segment in isolation}

The theory of wave power absorption is concerned with the waves generated by the WEC. For a single segment of
Pelamis in isolation, these may readily be found with a 3-D diffraction program. Figure (11) below shows a surface element mesh generated by the WS Atkins program AQWA-LINE (which gives substantially identical results to MITís program
WAMIT) \cite{Newm1998}. The segment is 3.5m diameter, 30m long with uniform density, and floating freely exactly half
immersed in water of infinite depth (and density $\rho$ = 1.000 tonne/m$^{3}$ with g = 10.00 m/s$^{2}$). \cite{Rainey}\

any waves propagating away to infinity by the complex function $A(\theta)$, from which the elevation of the waves at great
distance is defined as:

\begin{equation}
\
R\left\{\frac{A\left(\theta{}\right)ae^{i\left(kR-\omega{}t\right)}}{\sqrt{2\pi{}kR}}\right\}
\
\end{equation}

in the polar coordinates (R, $\theta$) these waves will ultimately decay inversely as the square root the distance
\textit{R} from the centre of the device, by energy conservation. The incident waves are assumed to travel in the direction $\theta$ = 0
in the sense of increasing \textit{R}, and to have angular frequency $\omega$ and wave number \textit{k}. Finally \textit{a} is the complex amplitude of the motion responsible for the waves of Equation (23). Thus when we consider free-floating behaviour in incident waves,\

\begin{equation}
\
\
R\left\{ae^{i\omega{}t}\right\}
\
\
\end{equation}

is the elevation of the incident waves at the centre of the segment. And when we consider heave motion in still water Equation (24) is the heave motion (positive upwards).
Figure (11) gives results (diffracted and diffracted+radiated waves) with the segment in incident waves 150m long (\textit{i.e}. as long as all five segments of \textit{Pelamis} combined) and also for heave motion in still water, at the same frequency. \cite{Rainey}\

Theoretically \cite{Newm1976}, A($\theta$)a for heave motion is equal to \

\begin{equation}
\
-i\omega{}H\left(\theta{}\right)e^{\left(i\pi{}/4\right)}/g
\
\end{equation}

where H($\theta$) is the\textit{ Kochin function} \cite{Newm1976}. In our case of long waves H($\theta$) may be approximated without difficulty as\

\begin{equation}
\
-k(-iωa)B
\
\end{equation}

 where \textit{B} is the waterplane area of the segment. This leads to a value of\ 

\begin{equation}
\
A\left(\theta{}\right)=k^2Be^{\left(i\pi{}/4\right)}
\
\end{equation}

which is shown in Figure (23), and agrees correctly with the computations. The more sophisticated slender-body approximation there will in fact recover the $\theta$- dependence seen in Figure (11) \cite{Newm1980}. For present purposes it is sufficient to observe that this dependence is small, and that the amplitude of the waves radiated by heave motion is proportional to the waterplane area \textit{B}. \cite{Rainey}\

\subsection{Energy flux in the far field: single segment}

Hitherto the segment has been floating freely; we are now in a position to assess the effect of incorporating a power
take-off so as to extract some wave energy. We will depart from the earlier literature cited above by considering the
energy flux in the far field directly. Since the freely-floating segment is practically transparent to the waves, and the only
significant wave radiation is from heave motion, we will take the far field waves produced by the segment (the
ëproduced wavesí) as being simply those due to the additional heave motion compared with the freely-floating segment.
Thus in Equation (23) we will take Equation (24) is this additional heave motion (positive upwards), and denote the incident waveelevation at the centre of the segment as\

\begin{equation}
\
R\left\{be^{i\omega{}t}\right\}
\
\end{equation}

The mean energy flux in the far field is the product of the wave pressure and the wave velocity, integrated over a fixed
cylindrical control surface where \textit{R} has some large value \textit{R}$_{c}$, say. The product of wave velocity and hydrostatic pressure
does not count because it is purely oscillatory, and the flux of kinetic energy does not count because it is of higher order
in wave height. The mean value of the product of any two complex oscillatory variables \textit{a} and \textit{b} is ½ℜ(āb), so if the incident and produced complex wave pressures are \textit{p}$_{I}$ and \textit{p}$_{P}$, and their complex velocities into the control surface are \textit{v}$_{I}$ and \textit{v}$_{P}$, the mean energy flux into the control surface is:

\begin{equation}
\
\int_{-\pi{}}^{\pi{}}\frac{1}{2}\
R\left\{\left(\overline{p_I+p_p}\right)\left({\upsilon{}}_I+{\upsilon{}}_p\right)\
\right\}R_c\frac{d\theta{}}{2k}
\
\end{equation}

Dealing with energy loss term first, and taking Equation (27) as above, this is readily integrated as:\

\begin{equation}
\
\int_{-\pi{}}^{\pi{}}\frac{1}{2}\
R\left\{\left\{\left(\overline{\rho{}g\frac{k^2Bae^{i\left(kR_c+\pi{}/4\right)}}{\sqrt{2\pi{}kR}}}\right)\left(-\omega{}\frac{k^2Bae^{i\left(kR_c+\pi{}/4\right)}}{\sqrt{2\pi{}kR}}\right)\right\}\
\right\}R_c\frac{d\theta{}}{2k}=
\
\end{equation}

\begin{equation}
\
\
=\int_{-\pi{}}^{\pi{}}\frac{\rho{}g\omega{}k^2B^2{\left\vert{}a\right\vert{}}^2}{8\pi{}}d\theta{}=-\frac{\rho{}g\omega{}k^2B^2{\left\vert{}a\right\vert{}}^2}{4}
\
\
\end{equation}

From an engineering point of view, however, it is very difficult to make such a device which is also capable of surviving extreme storms. \cite{Rainey, Lamb1932} \

The system is as efficient energy production. The engineering wave reflection absorption should be improved, so that the system support violent storms at sea.\

\subsection{Pelamis Wave Power}

Founded in 1998, Edinburgh-based Pelamis Wave Power (PWP) manufactures and operates Pelamis wave energy converters.\

In 2004, Pelamis Wave Power demonstrated their first full-scale prototype, the P1, at EMEC’s wave test site at Billia Croo. Here, the P1 became the world’s first offshore wave power converter to successfully generate electricity into a national grid. The device was 120m long, 3.5m in diameter and comprised four tube sections.\

The findings from Pelamis’ testing at EMEC between 2004 – 2007 led to the development of their second generation device – the P2. The P2 comprises five connected sections which flex and bend in the waves. This movement is harnessed by hydraulic rams at the joints which in turn drive electrical generators located inside the device. The device is 180m long, four metres in diameter and weighs approximately 1,350 tonnes.\

The first P2 machine, P2-001, was ordered by E.ON UK in 2009: the world’s first wave power machine to be purchased by a utility company. Arriving in Orkney in July 2010, the 750kW P2 Machine was successfully installed at the Billia Croo wave test site for the first time in October 2010. Following a three-year testing programme, the P2-001 has now returned to the ownership of Pelamis Wave Power, for continued demonstration alongside the ScottishPower Renewables owned P2-002.\

The testing or ‘work-up’ programme deployed by Pelamis is structured through a series of weather states, each with progressively higher wave heights. The P2 will be tested over a defined period of time in each state before graduating to the next. This approach allows progressive management of risk for the technology and the ability to find and handle any unexpected technical issues as they arise. Inspection and maintenance work is carried out at Lyness, where the machine is located when not at the wave test site, ready for redeployment in suitable weather windows.\

Whilst the P2 machines remain in Orkney, in November 2014 Pelamis Wave Power appointed an administrator to assess the options for future development. \cite{EMEC}\

\begin{figure}
	\begin{center}
		\includegraphics[scale=0.37]{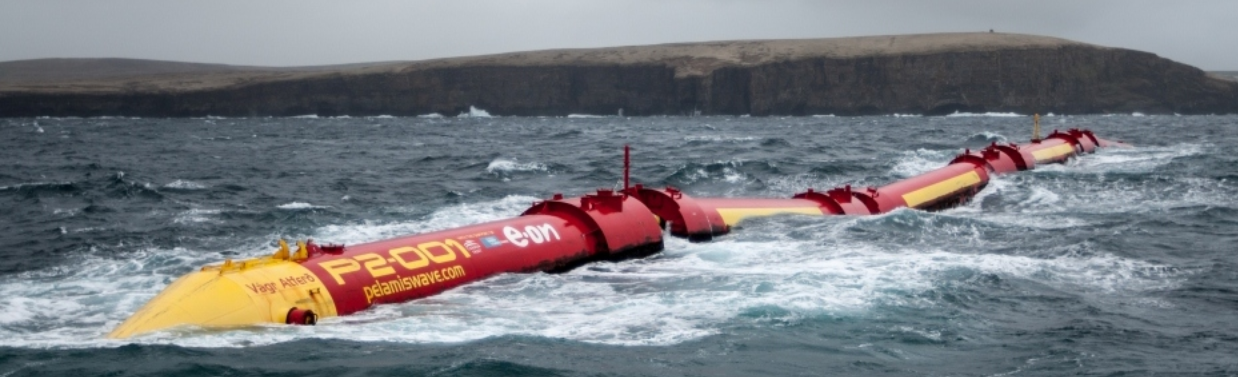}
		\caption{\small{The first P2 machine, P2-001, was ordered by E.ON UK in 2009: the world’s first wave power machine to be purchased by a utility company. Arriving in Orkney in July 2010,  the 750kW P2 Machine was successfully installed at the Billia Croo wave test site for the first time in October 2010. \cite{EMEC}}}\label{fig:sweeps}
	\end{center}
\end{figure}

\subsection{ScottishPower Renewables}
ScottishPower Renewables is part of Iberdrola, the world’s largest wind energy developer, with an operating portfolio of over 14,000 megawatts (MW) (as of March 2012). ScottishPower Renewables is responsible for progressing Iberdrola’s onshore wind and marine energy projects in the UK and Ireland, and offshore windfarms throughout the world, managing the development, construction and operation of all (current and potential) projects.\

In May 2012, the company’s second-generation Pelamis Wave Power device – the P2 – was deployed at EMEC’s wave test site for the first time on an adjacent berth to E.ON’s P2 device, as part of a unique joint working arrangement between the two renewable energy developers to maximise learning from operating and maintaining the machines as a wave farm. The machine is based at Lyness when not deployed on site.\

The 750kW P2 comprises five connected sections which flex and bend in the waves. This movement is harnessed by hydraulic rams at the joints, which in turn drive electrical generators located inside the device. The device is 180 metres long, four metres in diameter and weighs approximately 1350 tonnes.\

The experience gained from this project will play a vital role in ScottishPower Renewables’ plans to install 66 Pelamis machines in a 50MW project off Marwick Head in Orkney (to the north west of the Billia Croo wave test site), for which an agreement for lease has been awarded by the Crown Estate.\

Whilst the P2 machines remain in Orkney, in November 2014 Pelamis Wave Power appointed an administrator to assess the options for future development. \cite{EMEC1} \

\section{Conditions of the sea}
For a correct study of how efficient the Pelamis is, the tests must be carried out under conditions where the sea is stable, that is, its tide has no major variations. Thus, tests performed with the Pelamis P1, show that all four structures have different swings where the structure 1 acts as a "breakwater", causing it to pass right through the wave oscillating only a few degrees (approximately -6.1$^{\circ}$ to 5.2$^{\circ}$ in winter and -5.8$^{\circ}$ to 5.4$^{\circ}$ in summer). Already the structures 2 and 3, which are among the other two, oscillate more (-11$^{\circ}$ to 11.3$^{\circ}$ and -13$^{\circ}$ to 10$^{\circ}$ respectively), causing them to produce more energy, but the difference of tides generated by stations year generates an abnormal oscillation of structures 2, 3 and 4, which causes the energy output is not as efficient at certain times of the year, Figure (18). \cite{HeHongzhous2013, QINGuodong2013}

\section{Global Distribution of Annual Wave Power Flux from Wave Watch III Wind-Wave Model}
Generating energy from the natural movement of ocean waves through wave energy conversion (WEC) devices is
a promising technology in the early stages of development. Intended for energy planners and project developers,
this chapter of the report focuses on the potential of using WEC devices in developing member countries (DMCs)
of the Asian Development Bank (ADB).\

\begin{figure}
	\begin{center}
		\includegraphics[scale=0.39]{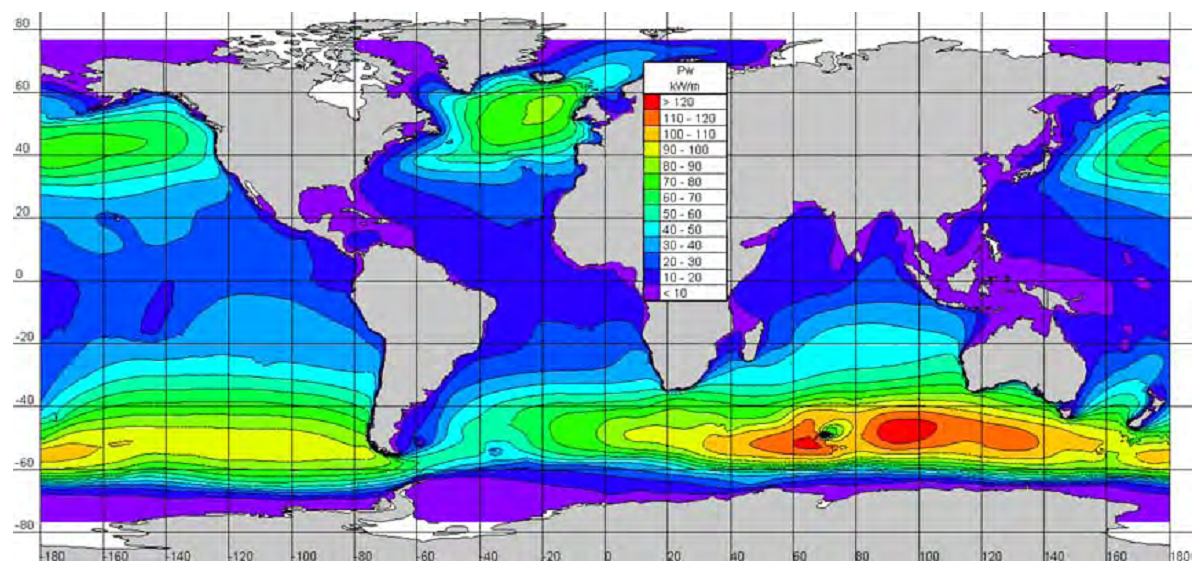}
		\caption{\small{Global Distribution of Annual Wave Power Flux from Wave Watch III Wind-Wave Model (kilowatt per meter). kW = kilowatt, m = meter, Pw = power flux.	Note: From 1997-2006 wind records, with a 0.5$^{\circ}$ latitude and longitude grid. Model calibrated with satellite altimeter data and buoy data from A. M. Cornett, 2008. A Global Wave Energy Resource Assessment. Proceedings of the 18th International Offshore and Polar Engineering Conference. Vancouver. 6–11 July. \cite{AsianDevelopmentBank2014}}}\label{fig:sweeps}
	\end{center}
\end{figure}

Currently, only one WEC device in the world has been transmitting electricity to distribution lines for more than
1 year: a 500-kilowatt (kW), shore-based, oscillating water column (OWC), land-installed marine-powered
energy transformer in Islay, Scotland, operational since 2000. In addition, some prototypes are being tested at the
European Marine Energy Centre and in Australia. Other than these early trials, however, the technology remains
largely experimental in nature.\

Numerous WEC concepts are discussed in the literature, ranging from simple sketches to reports of at-sea tests.
Some are shoreline-based, while others are seabed-mounted or moored in depths of less than 80 meters (m).
According to their directional characteristics, they can be classified as point absorbers, terminators, and
attenuators. Point absorbers have dimensions that are small relative to ocean wave lengths and are usually axissymmetric.\\

\textbf{1}. The principal axis of terminators is aligned perpendicularly to the direction of wave propagation; for
attenuators.\\

\textbf{2}. It is parallel to the direction of propagation. These have dimensions in the order of the wave lengths.\\

Given that the majority of DMCs that may have the appropriate resources for WEC are Pacific island DMCs, WEC devices currently being considered for Hawaii are assumed to be representative of future options for these DMCs. These can be categorized under two operating principles: wave-activated point absorbers and OWC. OWC devices use wave action to expand and compress air above a water column to rotate an air turbine generator (e.g., Oceanlinx). The wave-activated devices oscillate due to wave action relative to a fixed part of the device, and use one of three generation systems:\\

(i) a hydraulic system to turn a motor generator;\\

(ii) a linear generator, which generates electricity by moving a magnetic assembly within a coil;\\

(iii) or direct rack and pinion mechanical coupling. \cite{AsianDevelopmentBank2014}

\subsection{Wave Energy Resource}

Among different types of ocean waves, wind generated waves have the highest energy
concentration. Wind waves are derived from the winds as they blow across the oceans.
This energy transfer provides a natural storage of wind energy in the water near the free surface. Once created, wind waves can travel thousands of kilometres with little energy losses, unless they encounter head winds. Nearer the coastline the wave energy intensity decreases due to interaction with the seabed. Energy dissipation near shore can be compensated by natural phenomena as refraction or reflection, leading to energy
concentration (``hot spots").\

\begin{figure}
	\begin{center}
		\includegraphics[scale=0.38]{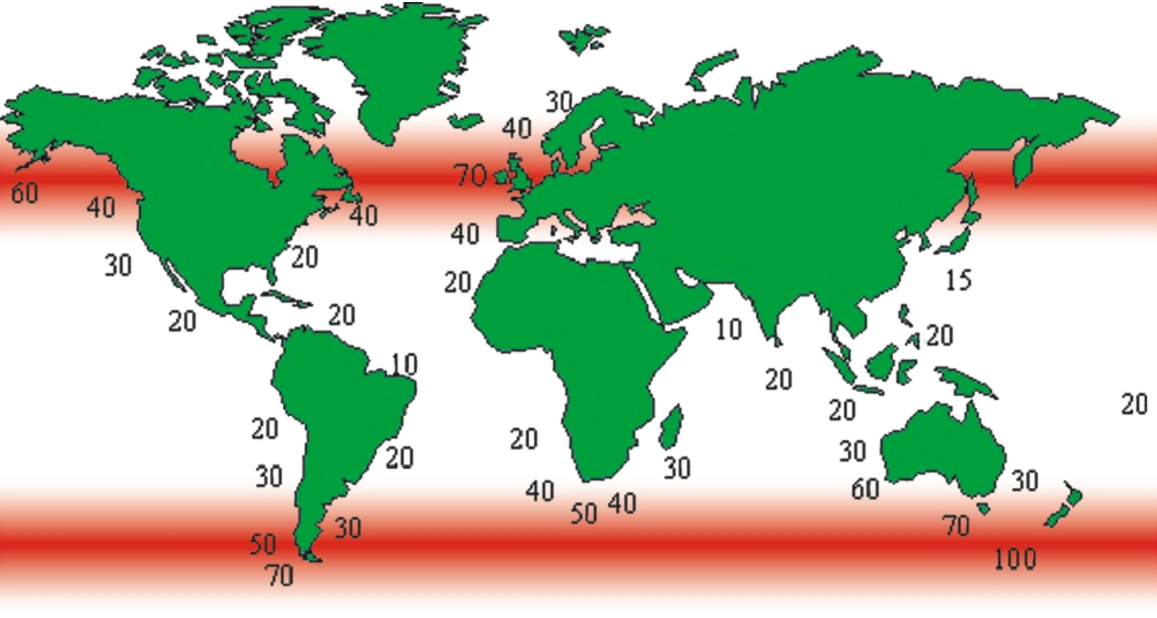}
		\caption{\small{Global wave power distribution in kW/m of crest length} \cite{EESD}}\label{fig:sweeps}
	\end{center}
\end{figure}

The power in a wave is proportional to the square of the amplitude and to the
period of the motion. Long period ($\sim$7-10 s), large amplitude ($\sim$2 m) waves have energy fluxes commonly exceeding 40-50 kW per meter width of oncoming wave. As most
forms of renewables, wave energy is unevenly distributed over the globe. Increased
wave activity is found between the latitudes of $\sim$30$^{\circ}$ and $\sim$60$^{\circ}$ on both hemispheres,
induced by the prevailing western winds (Westerlies) blowing in these regions.\ \cite{EESD}

\begin{figure}
	\begin{center}
		\includegraphics[scale=0.45]{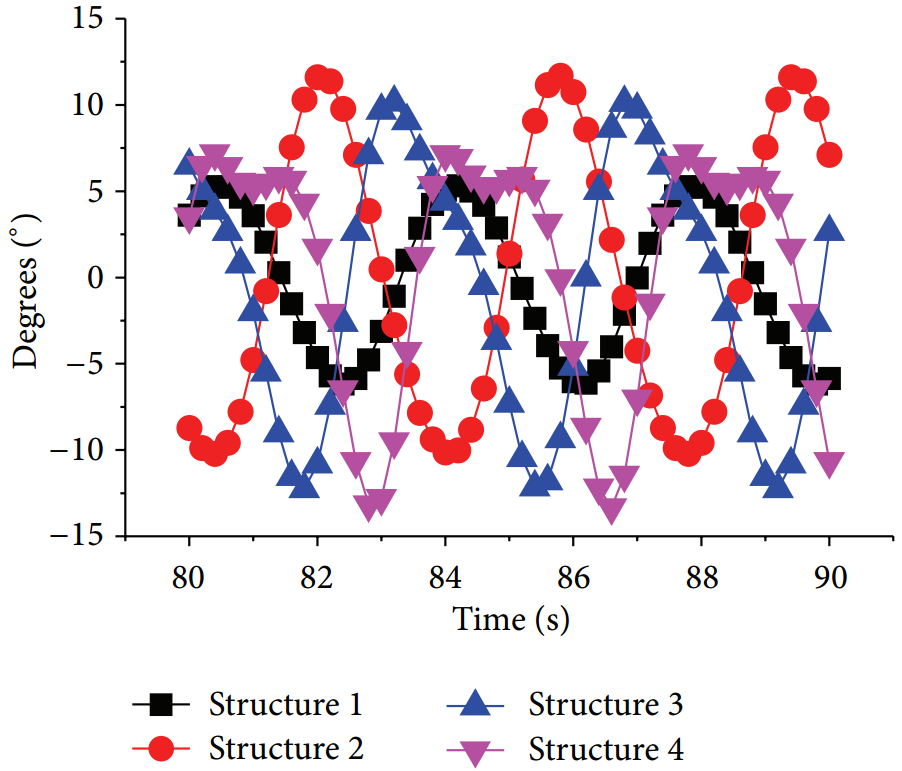}\\
		\includegraphics[scale=0.45]{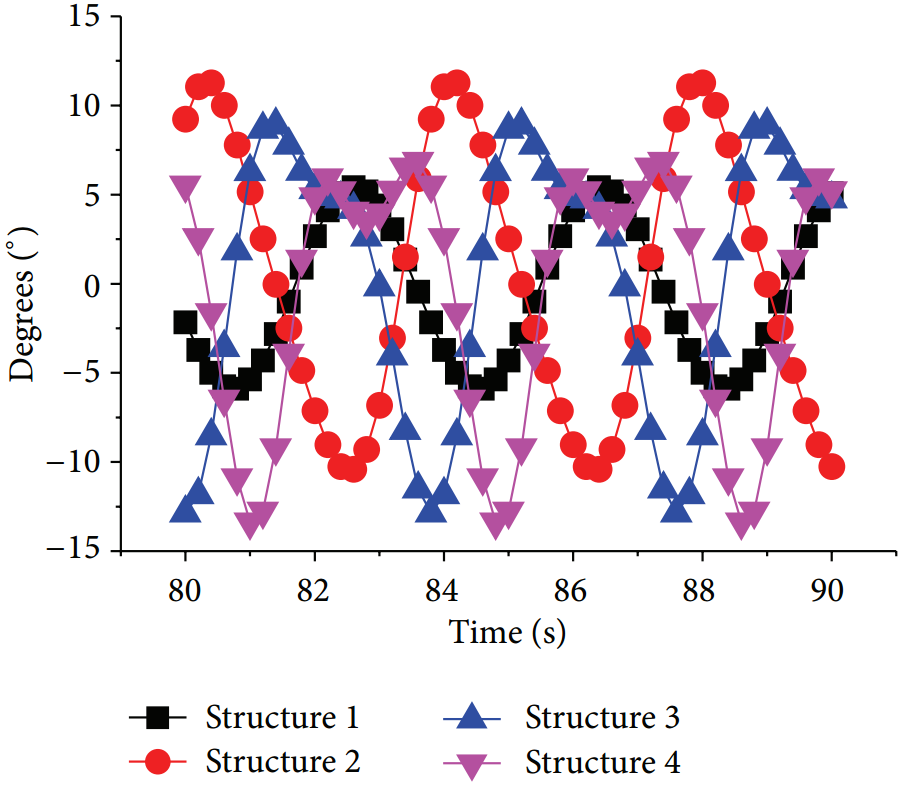}
		\caption{\small{Oscillation graph of Pelamis P1 structures. The four-section swing angles in spring (up) and summer (down). \cite{HeHongzhous2013}}}\label{fig:sweeps}
	\end{center}
\end{figure}

\section{Conclusions}
By analyzing the data we can understand that there are various forms of clean energy harvesting. Where in some of them has a higher cost, but the system studied, Pelamis, still under development, can not generate much energy, or have a very good cost benefit at the time, but it has enormous potential thanks to the use of tides that can be optimized so that their cost / benefit and production increase exponentially. Given the facts, it is safe to say that it is only a matter of time until he is one of the best and biggest sources of clean energy.

\bibliographystyle{unsrt}
\bibliography{journals}

\end{document}